# How to Survive a Learning Management System (LMS) Implementation?   A Stakeholder Analysis Approach

Ajayi Ekuase-Anwansedo
Science and Math Education
Southern University and A & M
Baton Rouge, Louisiana, USA
ajayi_anwansedo_00@subr.edu

Susannah F. Craig
Science and Math Education
Southern University and A & M
Baton Rouge, Louisiana, USA
susannah_craig@subr.edu

Jose Noguera
Management and Marketing
Southern University and A & M
Baton Rouge, Louisiana, USA
jose_noguera@subr.edu

## ABSTRACT
To survive a learning management system (LMS) implementation an understanding of the needs of the various stakeholders is necessary.  The goal of every LMS implementation is to ensure the use of the system by instructors and students to enhance teaching and communication thereby enhancing learning outcomes of the students. If the teachers and students do not use the system, the system is useless.  This research is motivated by the importance of identifying and understanding various stakeholders involved in the LMS implementation process in order to anticipate possible challenges and identify critical success factors essential for the effective implementation and adoption of a new LMS system.   To this end, we define the term stakeholder. We conducted a stakeholder analysis to identify the key stakeholders in an LMS implementation process. We then analyze their goals and needs, and how they collaborate in the implementation process. The findings of this work will provide institutions of higher learning an overview of the implementation process and useful insights into the needs of the stakeholders, which will in turn ensure an increase in the level of success achieved when implementing a LMS.

## 1 Introduction
Institutions of higher learning use Learning management systems (LMS) extensively.  Several studies demonstrate that the use of LMS is beneficial to effectively manage the classroom, circulate course materials and in communication and collaborative activities [23], [1].

Therefore, the goal of implementing a LMS in higher education is to ensure its use, to achieve these benefits.  To ensure the success of the implementation process all stakeholders must work together and be fully involved in the process [25].

However, this is not the case in the LMS implementation process. Typically, the university management, the chief information officer (CIO) and faculty representatives make the administrative decision to implement a new LMS or transition from one LMS to another [19] without any input from the students.

   Students are often not included in the decision-making process of an LMS implementation process even though the usefulness and the level of success of the LMS is determined in part by the frequency of student use and its impact on educational outcomes [6]. Furthermore, even though students are most frequently identified by other major stakeholders in institutions of higher learning as major stakeholders and they have direct impact on the University's income, [6], their contributions during the LMS implementation process is often ignored. An assumption is made that students will adopt whatever LMS is presented to them, ignoring the fact that they are tech savvy.

Some of the reasons for the change may include budget issues, cost of license of the LMS [17], [22] poor support system [17] and the popularity of the LMS [16].  After, the administrative decision is made, the chief information officer (CIO) and his team are saddled with the responsibility of implementing the technical aspects of the LMS process, developing training sessions and courses to introduce faculty, students and help desks personnel to the new LMS.  This process is depicted in  figure.1 below

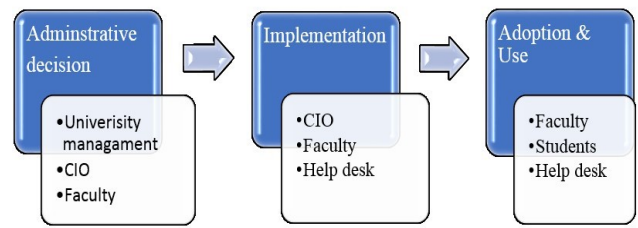

**Figure 1: LMS implementation process and accompanying stakeholders.**

Subsequently, a pilot of the LMS is conducted, to gradually introduce faculty and staff to the new LMS.  The goal of conducting the pilot of the system is to evaluate the usefulness and usability of the system, improve, and identify areas for improvement based on user response [2]. Therefore, it is cost effective to engage key stakeholders - faculty and students at the decision stages of the implementation process rather than at the pilot phase of the process.

Partial engagement or lack of engagement of the end-user stakeholders in the decision-making process leads to reluctance and limited use of the system by end users – faculty [17], [1] and students [17]. Some of the reasons given by faculty for limited acceptance of the LMS includes lack of financial compensation to migrate courses to new platform, limited time and lack of clarity on how to implement the new LMS [1], [20]. Students will only use the LMS in classroom during instruction, which defects one of the purpose of implementing an LMS – to enable students have access to course content and encourage learning outside the classroom [17].

## 2 Purpose of Study

The purpose of this study is to examine, using stakeholder analysis, the LMS implementation process to identify the key stakeholders and examine their interests and concerns. Stakeholder analysis approach (SA) involves the process of identifying the key players in the LMS implementation process and devising ways to manage them [24]. Stakeholder identification and involvement early in the implementation process increases the level of acceptance of the LMS and the likelihood of success of the implementation process [16[, [15].

Although stakeholder analysis has been widely explored in information systems and in other fields, SA in LMS implementation process are scant in literature [6]. Earlier research in the LMS process did not provide an all-inclusive and clear view of stakeholder's engagement in the LMS process.

Some research focused on faculty [21], [7] and others on students [26], [11] or both [12]. Furthermore, studies that attempt to examine all the stakeholders focus on their concerns and interests and did not utilize stakeholder analysis to comprehensively investigate their level of involvement in the LMS implementation process as well as their interaction with other stakeholders.

Consequently, our study investigates this gap and aims to provide a comprehensive view and a better understanding of stakeholder's involvement in LMS implementation process. The first part of this study which is covered in this paper comprises the identification of stakeholders and their roles in the LMS implementation process by reviewing and synthesizing existing literature. The second part of this study employs stakeholder analysis to comprehensively analyze the LMS implementation process to identify the interests and concerns of various stakeholders as well as their interactions using surveys and document analysis. **3 Stakeholder Analysis (SA)**

Stakeholder analysis has been defined is a process of identifying stakeholders and determining stakeholder's involvement in the decision-making process [19]. It is also seen as a tool or set of tools used to gather information about stakeholders to analyze their goals, interests and their relationships to estimate their level of impact in decision making or implementation processes [24]. Stakeholder analysis has been used in a number of studies to investigate the role of stakeholders in information systems project abandonment [16], in higher education research to explore valorization in higher education [3] to identify and prioritize stakeholders in a University [6], in project management to investigate stakeholder engagement [15].

### 3.1 Importance of Stakeholder Analysis in LMS Implementation Process

Stakeholder analysis in LMS implementation process contributes to the overall understanding of the needs, interest and communication requirements of stakeholders involved in the process. SA enable administrators determine the possibility of faculty and students adopting the LMS, anticipate areas of resistance to adoption, or other problem areas, establish lines of communication and allocate resources accordingly. SA also increases the chances of stakeholders adopting the new LMS.

The importance of identifying and understanding various stakeholders in the LMS implementation process especially the end user stakeholders is vital because there are distinct differences in faculty and student's adoption and interaction with the LMS [12]. In addition, the end user stakeholder's decision to adopt an LMS may differ based on several factors other than the quality, availability, and the price of the LMS. Environmental factors such as earthquakes and floods, [8], [13], personal factors such as habit [18] and financial factors such as faculty compensation [1], have been known to influence stakeholder's decision to adopt a LMS. In this study, we define stakeholder analysis as the process of identifying key stakeholders, classifying stakeholders based on their level of importance in the learning management implementation process and investigating interactions between them with the goal of managing each stakeholder accordingly. Thus, the stakeholder analysis of the LMS implementation process consist of the following steps:

i.) Identifying the key stakeholders ii.) Classifying stakeholders based on their level of importance.
iii.) Investigating interactions between stakeholders.

## 4 Identifying Stakeholders

### 4.1 Importance of Stakeholder Analysis in LMS Implementation Process

A stakeholder in an organization or a project is an individual or an organization or group who is affected or who has an interest in the organization's activities or in the project [17]. Similarly, the stakeholders in any LMS implementation process are those who are affected by the LMS implementation.

Table 1 below depicts stakeholders in LMS identified by various researchers

| Stakeholders | Researchers |
|---|---|
| Students, instructors, educational institutions, content providers, technology providers, accreditation providers and employers. | [25] |
| Instructors and students | [17] |
| Users, faculty, staff and administration | [20] |

**Table 1: Stakeholders in Learning management systems**

## 4.2 Types of Stakeholders in the LMS Implementation Process

In numerous studies, stakeholders are either internal or external stakeholders depending on if they are within or outside the institution. Primary or secondary stakeholders depending on their level of influence in the process.

This study classifies stakeholders in the LMS implementation process based on their current roles in the process:

I. Decision-making stakeholders
II. End-user stakeholders   III.   Vendor stakeholders
IV. Support stakeholders

The decision-making stakeholders includes entities who make the decision to introduce a new LMS or transition from one to the other. Some members of this group are also financially responsible for the entire LMS implementation process. This group is made up of the administration, the CIO and faculty representatives. The end-user stakeholders represent those who will eventually use the LMS. They are usually faculty and students. The vendor stakeholders are the creators or the sellers of the LMS. The support stakeholders consist of the support staff responsible for providing support for users using the system. Often these roles overlap, and a stakeholder takes on more than one role. For instance, the faculty is an end user stakeholder as well as decision making stakeholder and sometimes acts as a support stakeholder for the students. Figure 2 below attempts to depict these relationships in the process.

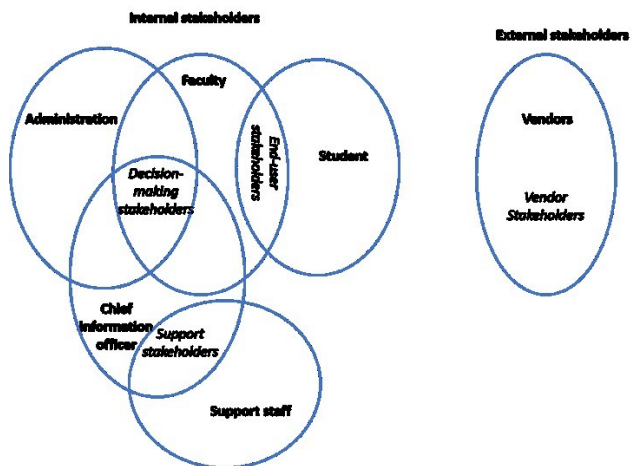

**Figure 2: Types of Stakeholders and their roles in the LMS implementation process**

## 5 Future Research

Future research will employ stakeholder analysis using a case study approach to investigate stakeholders in higher education. Questionnaires and interviews will be conducted with key stakeholders to determine their roles, level of importance and interactions in the LMS implementation process.

## CONCLUSION

The success of the LMS implementation process is determined by the level of adoption and use of the LMS by the end users – faculty and students. Stakeholder analysis presents a way to proactively engage and manage end users to ensure that the implementation process is successful. This paper examined the relevance of stakeholder analysis in the LMS implementation process.